\documentclass[sigconf,nonacm,authorversion]{acmart}

\usepackage{enumitem}
\usepackage{flushend}
\usepackage{subfiles}
\usepackage{enumitem}
\usepackage{color, colortbl}
\usepackage{xcolor}
\usepackage{tikz}
\usepackage{caption}
\usepackage{comment}
\usepackage{balance}
\usepackage{hyperref}       
\usepackage{url}            
\usepackage{booktabs}       
\usepackage{amsfonts}       
\usepackage{nicefrac}       
\usepackage{microtype}      
\usepackage{lipsum}
\usepackage{amsthm}
\usepackage{xspace}
\usepackage[normalem]{ulem}
\usepackage{algorithm}
\usepackage{algorithmic}
\usepackage{amsmath}
\usepackage{subcaption}
\usepackage{siunitx}        
\setlist{noitemsep}
\renewcommand{\paragraph}[1]{\noindent {\bf #1}}

\renewrobustcmd{\bfseries}{\fontseries{b}\selectfont}
\renewrobustcmd{\boldmath}{}
\newrobustcmd{\B}{\bfseries}

\definecolor{Gray}{gray}{0.9}

\newcolumntype{a}{>{\columncolor{Gray}}r}
\newcolumntype{b}{>{\columncolor{lightgray}}r}

\newcommand{\figref}[1]         {Figure~\ref{fig:#1}}

\newcommand{\secref}[1]         {Section~\ref{sec:#1}}

\newcommand{\secreftwo}[2]      {Sections \ref{sec:#1} and~\ref{sec:#2}}

\newcommand{\figreftwo}[2]      {Figures \ref{fig:#1} and~\ref{fig:#2}}

\renewcommand{\paragraph}[1]{\vspace{0.09in}\noindent{\bf \boldmath #1.}}

\newcommand{\numreorderings}{10\xspace}
\newcommand{\nummatrices}{110\xspace}
\newcommand{\numclusterings}{3\xspace}
\newcommand{\matrixA}{matrix $A$\xspace}
\newcommand{\matrixB}{matrix $B$\xspace}
\newcommand{\matrixC}{matrix $C$\xspace}

\newcommand{\MatrixB}{Matrix $B$\xspace}
\newcommand{\csrcluster}{\texttt{CSR\_Cluster}\xspace}
\newcommand{\meanhcspeedup}{1.39$\times$\xspace}
\newcommand{\rangehcspeedup}{0.96-1.75$\times$\xspace}
\newcommand{\hcoverheadfactor}{$20\times$\xspace}
\newcommand{\hcoverheadproblems}{$90\%$\xspace}
\newcommand{\percentinputshcspeedup}{70\%\xspace}

\definecolor{magenta4}{rgb}{0.5625,0,0.5625}
\definecolor{green4}{rgb}{0,0.5625,0}
\definecolor{orange4}{rgb}{0.98,0.31,0.09}

\newcommand{\helen}[1]{{ \textcolor{magenta4}{Helen: {#1}}}}

\begin{document}

\title{Improving SpGEMM Performance Through Matrix Reordering and Cluster-wise Computation}
\author{Abdullah Al Raqibul Islam}
\affiliation{
	\institution{University of North Carolina at Charlotte}
	\city{Charlotte}
    \state{NC}
    \country{USA}
}
\email{aislam6@charlotte.edu}

\author{Helen Xu}
\affiliation{
	\institution{Georgia Institute of Technology}
	\city{Atlanta}
    \state{GA}
    \country{USA}
}
\email{hxu615@gatech.edu}

\author{Dong Dai}
\affiliation{
	\institution{University of Delaware}
	\city{Newark}
    \state{DE}
    \country{USA}
}
\email{dai@udel.edu}

\author{Aydın Buluç}
\affiliation{
	\institution{Lawrence Berkeley National Lab}
	\city{Berkeley}
    \state{CA}
    \country{USA}
}
\email{abuluc@lbl.gov}

\begin{abstract}
  Sparse matrix-sparse matrix multiplication (SpGEMM) is a key kernel in many
  scientific applications and graph workloads. Unfortunately, SpGEMM is bottlenecked by data movement due to its irregular memory access patterns.
  Significant work has been devoted to developing row reordering schemes towards improving locality in sparse operations, but prior studies mostly focus on the case of sparse-matrix vector multiplication (SpMV).

  In this paper, we address these issues with \emph{hierarchical clustering} for
  SpGEMM that leverages both row reordering and cluster-wise computation to
  improve reuse in the second input (B) matrix with a novel row-clustered matrix
  format and access pattern in the first input (A) matrix. We find that
  hierarchical clustering can speed up SpGEMM by \meanhcspeedup on average 
  with low preprocessing cost (less
  than \hcoverheadfactor the cost of a single SpGEMM on about \hcoverheadproblems of
  inputs). Furthermore, we decouple the reordering algorithm from the clustered
  matrix format so they can be applied as independent optimizations.

Additionally, this paper sheds light on the role of both row reordering and clustering independently and together for SpGEMM with a comprehensive empirical study of the effect of \numreorderings different reordering
    algorithms and \numclusterings clustering schemes on SpGEMM performance on a suite of \nummatrices matrices. We find that reordering based on graph partitioning provides better
SpGEMM performance than existing alternatives at the cost of high preprocessing time. The evaluation demonstrates that the proposed hierarchical clustering method achieves greater average speedup compared to other reordering schemes with similar preprocessing times.



\end{abstract}

\maketitle

\thispagestyle{plain}
\pagestyle{plain}



\section{Introduction}
\label{sec:intro}

Sparse matrix-sparse matrix multipication (SpGEMM) is a key kernel in many
machine learning, matrix, tensor, and graph workloads. For example, it underlies
key algorithms in sparse deep neural
networks~\cite{han2015deep,parashar2017scnn}.  Numerical applications such as
the Algebraic Multigrid (AMG) method for solving sparse systems of linear
equations~\cite{ballard2016reducing}, volumetric mesh
processing~\cite{mueller2017ternary}, and simulation~\cite{canning1996n} use
SpGEMM as a subroutine. Finally,  
key graph analytics~\cite{kepner2015graphs}, such as breadth-first
search~\cite{gilbert2006high}, betweenness
centrality~\cite{bulucc2011combinatorial}, Markov
clustering~\cite{azad2018hipmcl}, label propagation~\cite{raghavan2007near}, triangle counting~\cite{azad2015parallel},
 peer-pressure clustering~\cite{ shah2007interactive}, and similarity
search~\cite{he2010parallel, agrawal2016exploiting} can be expressed as SpGEMM.

\begin{figure}[]
    \centering
    \includegraphics[width=1\linewidth,page=1]{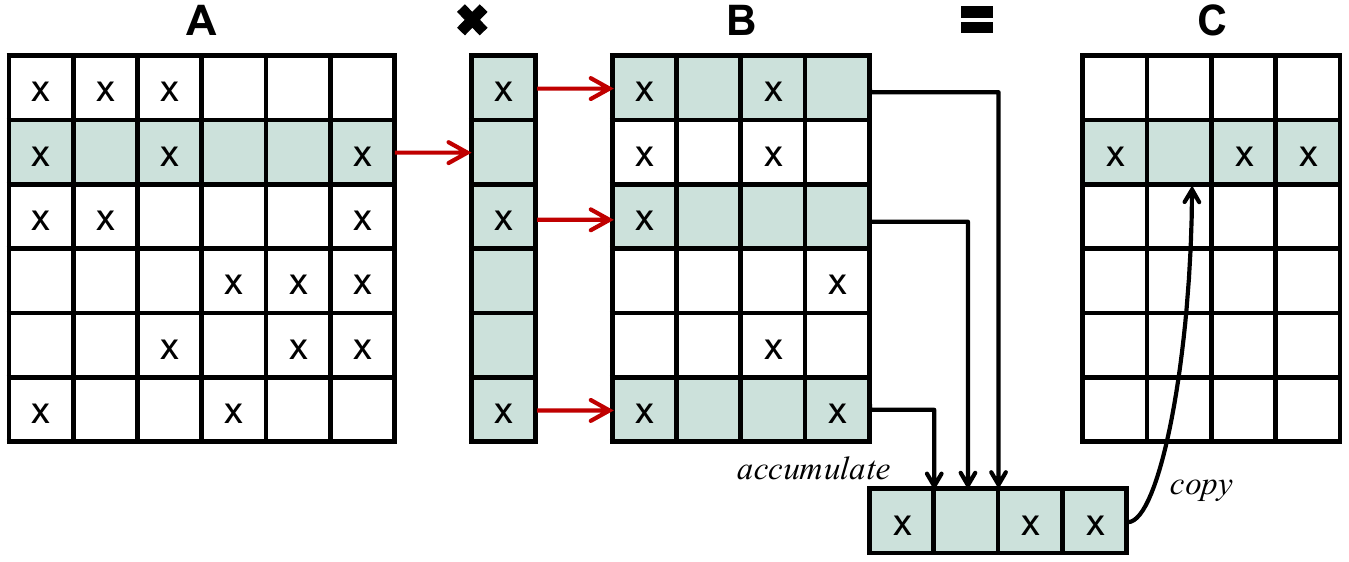}
    \caption{Basic SpGEMM on row order (Gustavson,
      1978)~\cite{gustavson1978two}.}
    \label{fig:gustavson-spgemm}
    \vspace{.5cm}
\end{figure}

SpGEMM is bottlenecked by \emph{memory traffic} and \emph{data movement} (i.e.,
it is memory-bound) due to its irregular access pattern~\cite{gamma}. As shown
in~\figref{gustavson-spgemm}, Gustavson’s algorithm, the go-to method for
multiplying sparse matrices, runs over each row (or column) of the first matrix and accumulates intermediate
products over a workspace that ultimately becomes the output row (or column)~\cite{gustavson1978two}. This algorithm raises two critical types of
memory access challenges: (1) irregular accesses to the second \emph{input
  matrix}, 
and (2) irregular accesses to the intermediate \emph{sparse
  accumulator}~\cite{gilbert1992sparse}, which stores the sparsity structure and
results of each output row. In this paper, given an SpGEMM, we will refer to the
first input matrix as $A$, the second input matrix as $B$, and the output as
matrix $C$ (i.e., $A \times B = C$).

\paragraph{Hierarchical sparse matrix reordering via clustering}
To address the performance impact of irregular memory accesses to the second \emph{input matrix}, prior work introduced a row reordering technique based on
\emph{hierarchical clustering}~\cite{JiangHoAg20}, which can improve the
performance of tiled sparse-dense matrix multiplication (SpMM) and sampled
dense-dense matrix multiplication (SDDMM) by over $3\times$ in some cases.
Hierarchical clustering implicitly defines a row ordering by finding clusters of
similar rows with locality-sensitive hashing and then reordering rows
accordingly for improved locality. Their high-level algorithm
first (1) defines groups of similar rows (in this instance, through
locality-sensitive hashing (LSH)~\cite{leskovec2020mining}), and then (2)
reorders the matrix based on the clustering to bring similar rows close
together.

This initial formulation of hierarchical clustering demonstrates the potential
for reordering in sparse BLAS-3 (matrix-matrix) operations with one sparse operand (out of three possible: two inputs and one output). However, their hierarchical clustering has three
major drawbacks: (1) the preprocessing time takes several orders of magnitude
greater than the actual kernel time, (2) the applicability of their method to sparse BLAS-3 operations with more than one sparse operand is unknown, and (3) the reordering leaves performance
on the table by storing the clustered matrix in row-major order in the classical
Compressed Sparse Row (CSR) format~\cite{TinneyWa67}. LSH is known to be
computationally expensive relative to a single sparse matrix
operation. Furthermore, the reordered matrix is still stored in row-major order
in CSR, so even if similar rows are grouped together, the relevant rows in the
B matrix may be evicted when moving between rows.

\paragraph{Improving hierarchical clustering} We introduce a novel formulation
of \emph{hierarchical clustering} for SpGEMM that resolves these drawbacks with
(1) faster identification of similar rows via SpGEMM and (2) a sparse matrix
format and dataflow for storing and processing similar rows. As we shall see,
rather than using an expensive LSH computation, we can generate similar row
pairs as candidates for clusters with a single SpGEMM between a matrix $A$ and
its transpose $A^T$. Furthermore, to capture the resulting cluster structure, we introduce a new sparse matrix format called ``\csrcluster'' that supports efficient
column-wise processing in clusters of $A$, improving reuse in accesses to rows
of $B$. We find that hierarchical clustering can speed up SpGEMM by \meanhcspeedup on average and up to 4.68$\times$  (between \rangehcspeedup on most inputs) with low preprocessing cost (less
  than \hcoverheadfactor the cost of a single SpGEMM on about \hcoverheadproblems of inputs). 

At a high level, the proposed SpGEMM can be viewed as a row reordering algorithm
with a corresponding format and access pattern change to take advantage of the
similarity of close rows. Next, we observe that the clustered formats could be
applied downstream of \emph{any row reordering} algorithm to potentially further
improve locality of reference in the $B$ matrix.

\paragraph{Sparse matrix reordering}
Significant effort has been devoted to developing row \emph{reordering}
algorithms~\cite{amestoy2004algorithm, cuthill1969reducing, george1973nested,
  george1989evolution, gilbert1986analysis, liu1976comparative,
  zhao2020exploring, balaji2023community, ChenSuZh25} for sparse matrices
towards improving locality of access, but the results are often mixed depending
on the input. Furthermore, many studies only test a small number of matrices and
demonstrate only about $10\%$ improvement in downstream sparse matrix-vector
multiplication (SpMV) performance ~\cite{gkountouvas2013improving,
  pichel2008reordering, pinar1999improving}. Finally, reordering schemes may
take orders of magnitude longer (e.g., $100\times$ or more) than the sparse
kernel time.

In this paper, we also present a comprehensive empirical study of row reordering
techniques in the context of SpGEMM, revealing new insights on tradeoffs between
performance improvement and preprocessing time from traditional reordering
algorithms. So far, row reordering has mostly been studied in the context of
SpMV, so it is not yet clear how it can impact SpGEMM, where both input matrices are
sparse.  Specifically, SpGEMM raises additional challenges when
compared to SpMV due to (1) more complex control flow, and (2) irregular accesses to the sparse accumulator. 

\begin{figure}[]
	\centering
	\includegraphics[width=.8\linewidth]{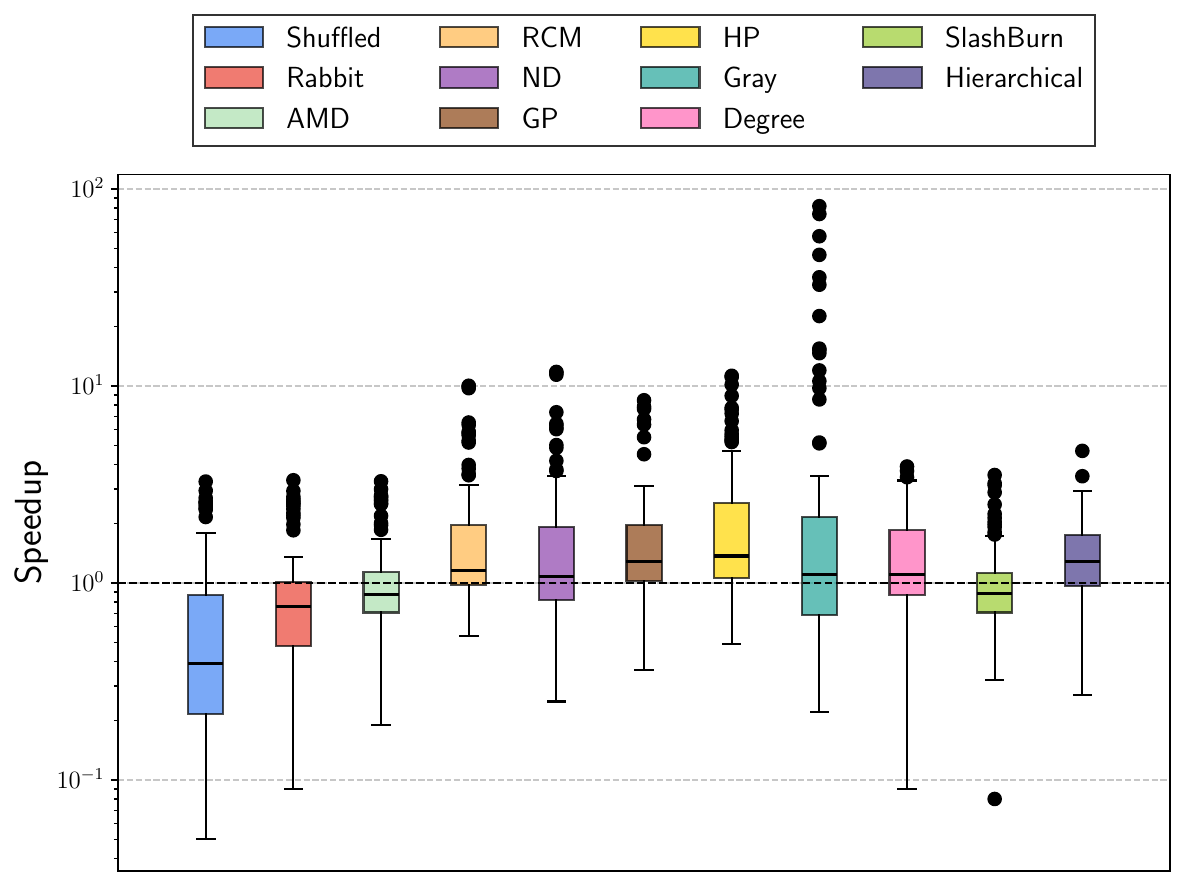}
    \caption{\small Speedup of row-wise SpGEMM after reordering, relative to the original matrix order.}
	\label{fig:perf_reorder_rowwise}
\end{figure}

\begin{figure}[]
	\centering
	\includegraphics[width=\linewidth]{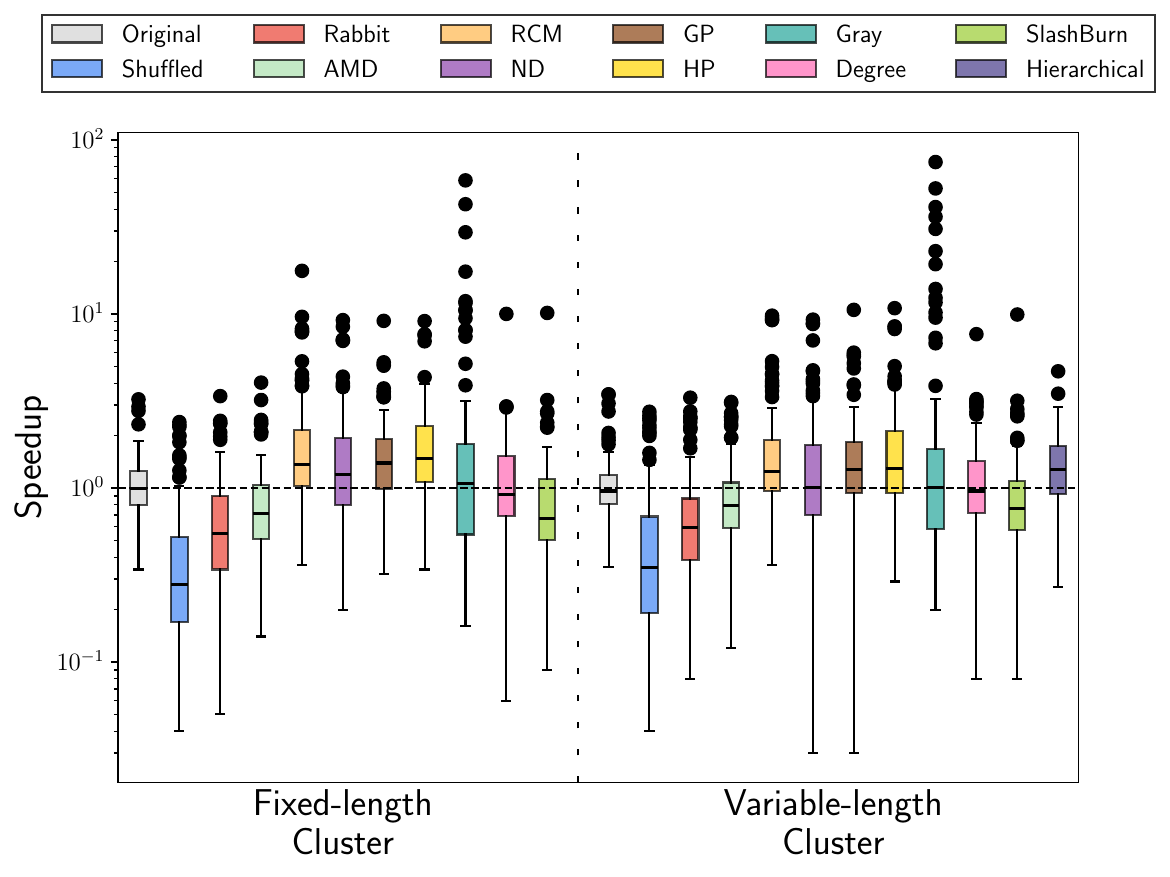}
    \caption{\small Cluster-wise SpGEMM with reordering, relative to row-wise SpGEMM performance on original order.}

\vspace{.5cm}
\label{fig:perf_reorder_cluster_base_rowwise}
\end{figure}

\paragraph{Comprehensively characterizing the impact of reordering and clustering}
In this paper, we introduce \emph{hierarchical clustering}, which leverages both
reordering and clustering, and the decoupled \emph{cluster-wise computation},
which can be applied downstream of any reordering algorithm. We comprehensively
evaluate clustering and reordering separately and together to understand
their effects in the context of SpGEMM.
Without loss of generality, we focus on the row-wise SpGEMM where clustering and reordering are both applied to the $A$ matrix.

For cluster-wise computation, we study three schemes to combine the computation
of multiple rows of the $A$ matrix to improve locality of reference to the $B$
matrix. First, we use a simple fixed-length (in terms of rows) scheme as a
baseline to see how much performance is possible without further
optimization. Next, we consider \emph{variable-length clustering}, which is the
decoupled first step of hierarchical clustering of finding groups of similar
nearby rows, but does not necessarily reorder the rows, enabling exploration of
different reordering schemes beyond LSH in the general high-level algorithm
flow. Finally, we present \emph{hierarchical clustering}, which combines a novel
fast reordering scheme based on SpGEMM with variable-length clustering.

For matrix reordering, we compare \numreorderings common techniques for row
reordering included in a comprehensive study of reordering for
SpMV~\cite{trotter2023bringing}, including graph-based
methods~\cite{amestoy2004algorithm}, Reverse
Cuthill-McKee~\cite{cuthill1969reducing, liu1976comparative}, and Gray
ordering~\cite{zhao_2020_exploring}, among others. Furthermore, we combine these
reordering schemes with fixed-length and variable-length clustering to explore
further improvements to locality after reordering.

To our knowledge, this is the most comprehensive study of both reordering and
clustering combined to date, with \nummatrices large matrices, \numreorderings
reordering algorithms, and \numclusterings clustering strategies.  Our
evaluation tests two common workloads for SpGEMM: squaring a sparse matrix $A$
(i.e., $A^2$), and square times tall-skinny
matrix~\cite{nagasaka_2018_high-performance}.



\paragraph{Contributions} The main contributions of this paper are as follows:

\begin{itemize}[leftmargin=10pt, nosep]
\item We introduce a new hierarchical clustering-based algorithm for shared-memory parallel SpGEMM, which
  combines (1) a novel reordering scheme based on finding similar rows via
  SpGEMM, and (2) a new sparse matrix format to capture the resulting clustered matrix
  structure.
\item A comprehensive empirical evaluation of the impact of \numreorderings
  reordering algorithms alone (without downstream clustering) on \nummatrices
  matrices in the context of SpGEMM.
\item An extensive empirical evaluation of hierarchical clustering, reordering,
  and reordering with clustering in the context of SpGEMM. The results
  demonstrate that hierarchical clustering generates high-quality reorderings
  and performance improvements for SpGEMM with minimal preprocessing time. Furthermore, it characterizes which reordering algorithms can improve SpGEMM performance and when clustering (both with and without reordering) can accelerate SpGEMM.
\end{itemize}

\paragraph{Evaluation summary}
~\figreftwo{perf_reorder_rowwise}{perf_reorder_cluster_base_rowwise} illustrate
the high-level results of performing $A^2$ with reordering both without and with
clustering, respectively, on a suite of \nummatrices matrices.

The key findings in this paper are as follows:

\begin{itemize}[leftmargin=10pt, nosep]
\item Hierarchical clustering speeds up SpGEMM by \meanhcspeedup on average
  and improves performance on \percentinputshcspeedup of the inputs.
\item Reordering based on graph and hypergraph partitioning (GP and HP) offers
  the highest geomean (1.77$\times$) and most consistent (on about $80\%$ of the inputs) speedups for both $A^2$ and
  tall-skinny SpGEMM, but has high preprocessing overhead (with many instances taking over $100\times$ of the cost of a single SpGEMM).
\item In general, matrix reordering algorithms expose a tradeoff between preprocessing cost and SpGEMM improvement.
\item Even without reordering, fixed-length and variable-length clustering can improve performance in approximately $45\%$ and $40\%$ of cases, respectively, with minimal preprocessing overhead.
\item Applying reordering before fixed-length and variable-length clustering can improve performance over clustering alone in many cases. For example, applying HP as a preprocessing step before cluster formation boosts performance on approximately $80\%$ of inputs. 
\item Combining reordering and clustering does not always compose: applying both techniques can improve performance over either one alone in some cases, but may degrade performance in others.
\end{itemize}


\section{Background, Related Work, And Motivations}\label{sec:background}

This section gives background on Compressed Sparse Row (CSR), a classical sparse matrix storage format, and how Gustavson's row-wise algorithm for SpGEMM uses it for efficient row access. Finally, it overviews several matrix reordering algorithms for accelerating sparse computations. These concepts are necessary to understand the improvements and evaluations in later sections.

\begin{figure}[]
    \centering
    \includegraphics[width=0.8\linewidth,page=6]{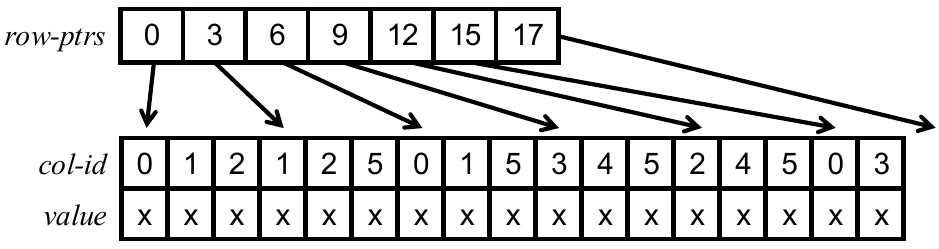}
    \caption{Sparse matrix of~\figref{gustavson-spgemm} in CSR format.}
    \vspace{.5cm}
    \label{fig:csr}
\end{figure}

\subsection{Sparse Matrix Storage Formats}
The Compressed Sparse Row (CSR) format~\cite{TinneyWa67} is the de facto standard for efficiently storing sparse matrices by eliminating zero entries. As illustrated in~\figref{csr}, CSR represents a sparse matrix using three arrays: \texttt{row-id}, \texttt{col-id}, and \texttt{value}. The \texttt{col-id} array stores the column indices of all non-zero elements, while the \texttt{value} array holds the corresponding non-zero values. The \texttt{row-id} array indicates the starting index in the \texttt{col-id} and \texttt{value} arrays for each matrix row. In the context of SpGEMM, this format enables efficient row-wise access and is particularly well-suited for implementations based on Gustavson’s algorithm.

\subsection{SpGEMM Kernel}
\figref{gustavson-spgemm} illustrates the core structure of Gustavson's SpGEMM algorithm~\cite{gustavson1978two}, one of the most widely adopted approaches. When matrices are stored in CSR format, the algorithm proceeds row-by-row over \matrixA (and correspondingly builds rows of the output matrix \matrixC). For each non-zero in a row of \matrixA, the corresponding row of \matrixB is accessed, and partial products are accumulated to compute entries in \matrixC.

Since the sparsity pattern and the number of non-zeros in matrix C are not known in advance, memory allocation is non-trivial. To address this, the algorithm performs an initial, lightweight traversal---known as the \textit{symbolic phase}---to count the number of non-zero entries and allocate space accordingly. This is followed by the \textit{numeric phase}, where the actual SpGEMM computation is performed.

During SpGEMM, the algorithm often maintains a sparse accumulator (referred to as \texttt{accumulate} in~\figref{gustavson-spgemm}) to collect intermediate products for each row. One common choice for the sparse accumulator is a hash table, due to its fast insertion and lookup capabilities~\cite{nagasaka_2018_high-performance}. After processing a row, the accumulated results are written back to the corresponding row in \matrixC (referred to as \textit{copy} in~\figref{gustavson-spgemm}), stored in CSR.

\subsection{Matrix Reordering}
A common strategy to optimize sparse matrix computations is matrix reordering, which improves data locality and computational efficiency. While reordering has been extensively studied in sparse kernels such as SpMM, SDDMM, and SpMV, to our knowledge, there has not yet been a comprehensive study of reordering for SpGEMM. In this study, we analyze the effect of \numreorderings reordering algorithms (listed in Table~\ref{tab:reorder-algos}) on SpGEMM performance across a diverse set of \nummatrices matrices.



\begin{table}[t]
    \centering
    \caption{\small Sparse matrix reordering algorithms used in this study.}
    \resizebox{\columnwidth}{!}{
        \label{tab:reorder_algorithms}
        \begin{tabular}{ll}
            \hline
            \textbf{Algo.} & \textbf{Description} \\
            \hline
            Original & Original input order \\
            Random & Random shuffle \\
            Rev. Cuthill–McKee (RCM)~\cite{liu1976comparative} & Bandwidth reduction via BFS \\
            Aprx. minimum degree (AMD)~\cite{amestoy2004algorithm} & Greedy strategy to reduce fill \\
            Nested dissection (ND)~\cite{george1973nested} & Recursive divide-and-conq. to reduce fill \\
            Graph partitioning (GP)~\cite{karypis1998fast} & METIS using edge-cut objective \\
            Hypergraph partitioning (HP)~\cite{catalyurek1999hypergraph} & PaToH using cut-net metric \\
            Gray code ordering~\cite{zhao2020exploring} & Splitting sparse and dense rows \\
            Rabbit order~\cite{arai2016rabbit} & Hierarchical community-based reordering \\
            Degree order & Reorder in descending order or degrees \\
            Slash-burn method (SB)~\cite{lim2014SlashBurn} & Recursively split rows into hubs and spokes \\
            \hline
        \end{tabular}
    }
    \label{tab:reorder-algos}
\end{table}

Many classical reorderings originate from sparse linear solvers. For instance, Cuthill-McKee (CM)~\cite{cuthill1969reducing} and its reverse variant RCM~\cite{George1979implementation, Gibbs1976algorithm} aim to reduce the matrix \textit{bandwidth}, which is defined as the maximum distance from the diagonal of any non-zero element in the matrix. Lower bandwidth generally leads to improved data locality during computation. Degree-based reorderings such as Minimum Degree and Approximate Minimum Degree (AMD)~\cite{amestoy2004algorithm, george1989evolution} aim to reduce \textit{fill-in}, which refers to the additional non-zero elements introduced during matrix factorization (e.g., LU or Cholesky) that were originally zero. These methods prioritize eliminating rows with fewer non-zeros to minimize fill-in, improving memory usage and computation time. Nested Dissection (ND)~\cite{gilbert1986analysis, george1973nested} recursively partitions the matrix using separators to reduce fill-in and improve parallelism, particularly effective for structured matrices.

We also evaluate graph-based reordering strategies motivated by their success in improving cache locality in graph analytics. Rabbit~\cite{arai2016rabbit} groups strongly connected nodes (communities) to enhance locality. Degree reordering packs high-degree vertices together to minimize cache line usage. SlashBurn~\cite{lim2014SlashBurn} recursively removes high-degree hubs to expose dense subgraphs and reorder them for better locality~\cite{Koohi2021exploiting}.

Additionally, we explore partitioning-based reorderings aimed at improving locality and parallel workload balance. Graph Partitioning (GP)~\cite{karypis1998fast} and Hypergraph Partitioning (HP)~\cite{catalyurek1999hypergraph} reorder matrix rows/columns based on partition assignments. We use METIS for GP (with an edge-cut objective) and PaToH for HP (using the cutpart objective and quality heuristics).

Gray code ordering~\cite{zhao2020exploring} arranges rows and columns using Gray code sequences, where consecutive indices differ by only one bit, helping to group structurally similar rows and improve data locality.

For evaluation, we adopt RCM and ND from Libmxt~\cite{libmtx}, Rabbit from Arai et al.~\cite{rabbitgit}, and the remaining implementations from SparCity~\cite{SparCity}. We also include a randomly shuffled ordering as an extreme baseline.

\section{Cluster-wise SpGEMM}\label{sec:clusterwise}
\begin{figure}[]
    \centering
    \includegraphics[width=1\linewidth,page=2]{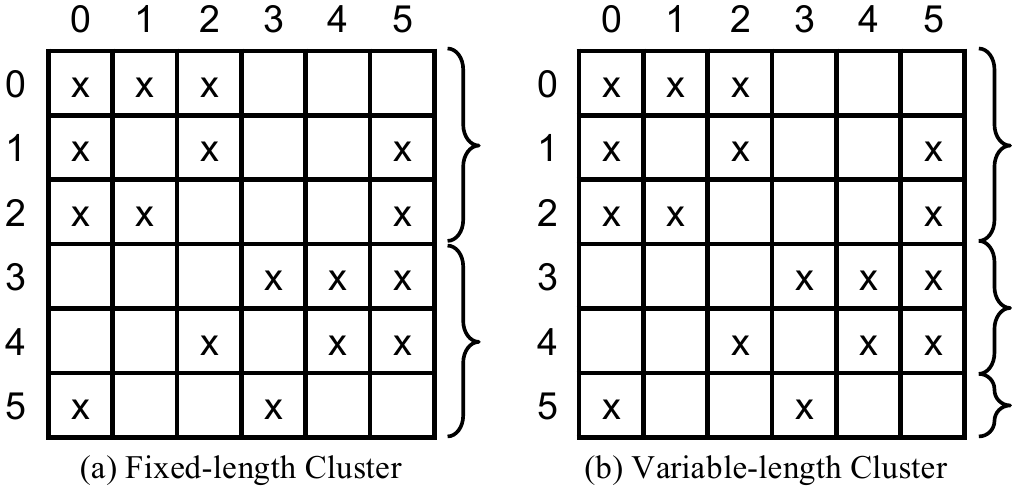}
    \caption{Clustering without changing order.}
    \label{fig:cluster_noorder}
\end{figure}

This section introduces hierarchical clustering, a novel optimization method for SpGEMM that combines both reordering and cluster-wise computation for improved locality and performance with low preprocessing overhead. First, this section will provide an example of potential reuse in SpGEMM by changing the access pattern in clusters. Along the way to hierarchical clustering, we decouple reordering and clustering and introduce two independent clustering schemes called ``fixed-length'' and ``variable-length'' clustering as a warmup that we will use later as independent optimizations that can be applied to any reordering scheme.

\paragraph{Motivation} 
Even if certain rows of \matrixB are accessed multiple times with potential for reuse, they may be evicted in traditional row-wise SpGEMM computation by the time they are requested again. 
First, each row of \matrixA may contain multiple non-zero elements, so, processing a single row of \matrixA can lead to accesses across multiple non-contiguous memory regions of \matrixB, increasing the likelihood of cache evictions.  
Second, consecutive rows in \matrixA may exhibit different sparsity patterns, with limited or no overlap in their non-zero column indices. For example, row 0 of \matrixB in~\figref{gustavson-spgemm} is accessed by rows 0, 1, 2, and 5 of \matrixA, as column 0 of \matrixA contains non-zero values in these rows, but row 0 of \matrixB may be evicted between processing rows of \matrixA depending on how many nonzeroes are in the rows. These challenges motivate a redesign of the traditional row- or column-wise SpGEMM computation strategy.

\begin{algorithm}[t]
\caption{ClusterWise\_SpGEMM}
\label{alg:cluster-spgemm}
\begin{algorithmic}[1]
\STATE // set matrix $C$ to $\emptyset$; \textcolor{blue}{$C$ in \csrcluster format}
\FORALL{$a_{i*}$ in matrix $A$ \textbf{in parallel}}
    \FORALL{$a_{ik}$ in \textcolor{blue}{cluster $a_{i*}$}}
        \FORALL{$b_{kj}$ in row $b_{k*}$}
            \FORALL{\textcolor{blue}{$a_{ikl}$ in col $a_{ik}$}}
                \STATE \textcolor{blue}{$c_{ijl} \gets c_{ijl} + a_{ikl} * b_{kj}$}
            \ENDFOR
        \ENDFOR
    \ENDFOR
\ENDFOR
\end{algorithmic}
\end{algorithm}

\subsection{Access Pattern and Matrix Format}
\label{sec:accesspattern}
To address these challenges, we decouple \emph{clustering} from \emph{reordering} to introduce \emph{cluster-wise SpGEMM}, which applies clustering to a sparse matrix with arbitrary (or even no) reordering. That is, we can first apply any reordering algorithm to a sparse matrix. As we shall see, there is no one-size-fits-all reordering method because the benefits of reordering depend on the matrix structure. After reordering, we can apply cluster-wise SpGEMM via ``fixed-length'' or ``variable-length'' clustering to improve reuse in \matrixB. 
Viewed in this way, the original formulation of hierarchical clustering~\cite{JiangHoAg20} is 
LSH-based reordering with ``variable-length clustering.'' 

The high-level idea behind cluster-wise SpGEMM is to enhance reuse of \matrixB row accesses across multiple \matrixA row computations by retaining relevant rows of \matrixB in the cache. Specifically, once a row of \matrixB is read, our goal is to keep it in the cache while processing several consecutive rows of \matrixA.



\paragraph{Access pattern} Algorithm~\ref{alg:cluster-spgemm} presents the cluster-wise SpGEMM algorithm, with the differences from the traditional row-wise Gustavson's algorithm 
highlighted in blue text.


In cluster-wise computation, we iterate over cluster IDs of \matrixA instead of row IDs, and traverse the non-zero column IDs of the merged rows (i.e., the cluster), as shown in Lines 2 and 3 of Algorithm~\ref{alg:cluster-spgemm}. \MatrixB rows are accessed in the same manner as in the row-wise SpGEMM computation. Once a row of \matrixB is accessed, the algorithm performs computations for all the rows within the corresponding \matrixA cluster (Line 5 of Algorithm~\ref{alg:cluster-spgemm}). This access pattern ensures that the relevant \matrixB row remains in the cache throughout the processing of multiple \matrixA rows, thus improving temporal locality.

For example, given the matrix in  \figref{gustavson-spgemm}, we would group the first three rows of \matrixA (as shown in~\figref{cluster_noorder}(a)) into a cluster, treating them as a unit of computation. Since clusters are processed column-wise, row 0 of \matrixB will be available in the cache while processing rows 0, 1, and 2 of \matrixA.

\begin{figure}[]
    \centering
    \includegraphics[width=1\linewidth,page=3]{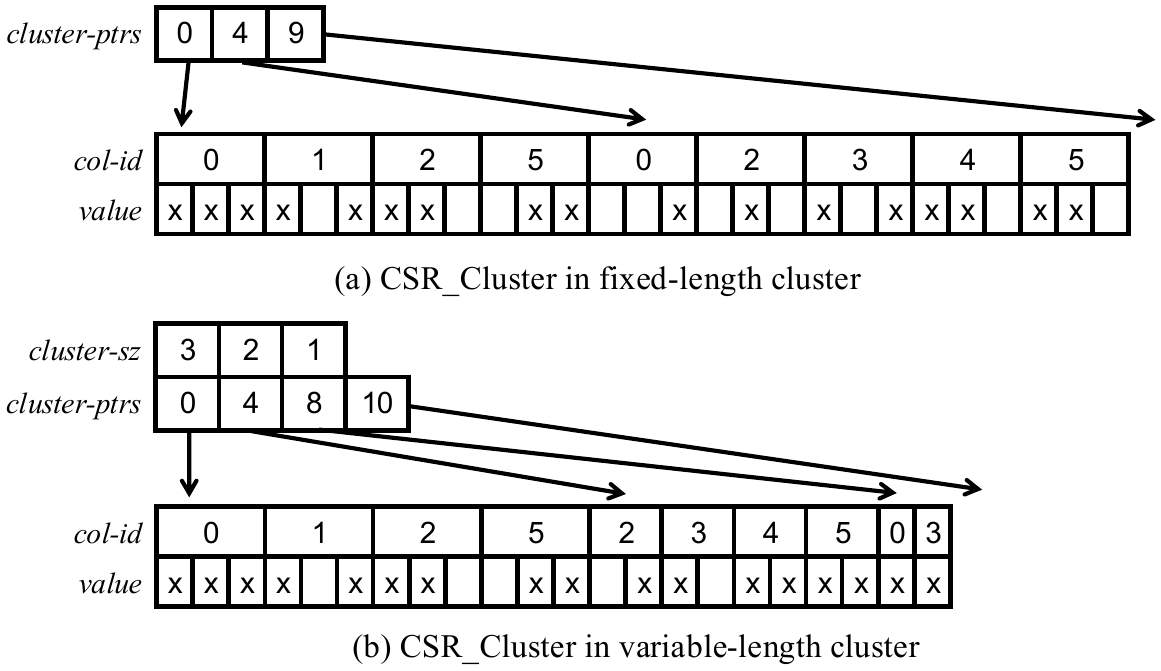}
    \caption{Sparse matrix of Fig.~\ref{fig:cluster_noorder} in \csrcluster format.}
    \label{fig:csr_cluster}
\end{figure}

\paragraph{Clustered matrix format} To efficiently support this computation pattern, we propose a new storage format called \csrcluster. \csrcluster groups multiple rows into clusters and stores their non-zero entries collectively by column, enabling efficient access patterns for cluster-wise SpGEMM. This layout improves temporal locality and cache reuse by increasing the likelihood that frequently accessed \matrixB rows remain in cache across several \matrixA row computations.~\figref{csr_cluster}(a) illustrates the \csrcluster representation for the cluster layout shown in~\figref{cluster_noorder}(a), assuming that two clusters are formed, each containing three consecutive rows.

Due to varying sparsity patterns, consecutive rows in a sparse matrix may not all have non-zero values in the same columns, leading to empty (or placeholder) positions in \csrcluster. For example, in \figref{cluster_noorder}(a), rows 0–3 all have nonzero entries in column 0, but while row 0 has no entry in column 5, rows 2 and 3 do. Consequently, when merging rows 0–3 into a cluster to construct \csrcluster, column 5 contains an empty (or placeholder) position—see cluster ID 0, column ID 5 in~\figref{csr_cluster}(a).~\secref{eval} fully characterizes the space overhead of \csrcluster, which is below $2\times$ for variable-length and fixed-length clustering in most (over 80\%) cases. 

In variable-length \csrcluster, the cluster sizes are stored in a separate array. This allows the row indices of the original matrix (as used in conventional CSR) to be derived implicitly from the cluster sizes, thereby eliminating the need to store them explicitly. In contrast, fixed-length \csrcluster does not require an additional array for cluster sizes, since all clusters are of uniform length. It is also worth noting that an additional array of pointers is required in variable-length \csrcluster to enable efficient access to the value array (not shown in~\figref{csr_cluster}(b)).

So far, we have explained how cluster-wise SpGEMM can improve cache locality by reusing \matrixB rows across multiple \matrixA row computations. However, to maximize the benefits of this approach, it is crucial to form clusters in a way that maximizes reuse. If consecutive rows in \matrixA do not share non-zero columns, then cluster-wise computation will incur overhead during iterating over columns in \csrcluster (as in Line 5 of Algorithm~\ref{alg:cluster-spgemm}). To address this, this section introduces three clustering strategies. 

\subsection{Clustering Without Changing Order}
In this section, we discuss cluster construction strategies independent of row reordering.
Specifically, we propose two straightforward strategies for cluster formation: (a) fixed-length clusters, where rows are grouped into clusters of equal size regardless of content, and (b) variable-length clusters, where the cluster size may vary depending on structural characteristics, but without relying on reordering rows of the sparse matrix. Either method can be applied to any matrix after any reordering scheme (or no reordering).

\paragraph{Fixed-length Clusters}
Many real-world sparse matrices from different scientific domains and optimization problems exhibit specific sparsity patterns, reflecting the structure of the underlying problem. One common example is a dense diagonal block pattern embedded within a sparse matrix~\cite{bates-matrix, matrix-dense-diagonal-block-atandt, matrix-dense-diagonal-block-gupta, matrix-dense-diagonal-block-trefethen}. 

When such patterns can be identified, the simplest and most lightweight approach for building clusters for SpGEMM is to form \emph{fixed-length clusters}, which groups an equal number of consecutive rows into each cluster. This method incurs minimal preprocessing overhead and aligns well with the block structure of the matrix. The number of rows per cluster may vary across matrices, depending on the structure of the diagonal blocks.

\figref{cluster_noorder}(a) illustrates an example of fixed-length clustering with clusters of three consecutive rows each, and ~\figref{csr_cluster}(a) provides the  corresponding {\csrcluster} representation.


\begin{algorithm}[t]
    \caption{Construct Variable-length Cluster}
    \label{alg:vlength-cluster}
    \begin{algorithmic}[1]
        \REQUIRE $A\_CSR[M][N], jacc\_th, max\_cluster\_th$
        \ENSURE $A\_CSR\_CLUSTER[M^*][N][K]$

        \STATE $clusters \leftarrow \text{Map}()$
        \STATE $rep\_row\_id, cluster\_id \leftarrow 0$
        \STATE $clusters[cluster\_id].\text{insert}(0)$; $cluster\_sz \leftarrow 1$;
        \FOR{$i = 1$ \TO $A.nrows - 1$}
            \STATE $j\_score \leftarrow A\_CSR.jaccard\_similarity(rep\_row\_id, i)$
            \IF{$j\_score < jacc\_th$ \OR $cluster\_sz = max\_cluster\_th$}
                \STATE $cluster\_id \leftarrow cluster\_id + 1$;
                \STATE $rep\_row\_id \leftarrow i$; $cluster\_sz \leftarrow 1$
            \ENDIF
            \STATE $clusters[cluster\_id].\text{insert}(i)$
        \ENDFOR
        \STATE

        \STATE $A\_CSR\_CLUSTER(A\_CSR, clusters)$
        \RETURN $A\_CSR\_CLUSTER$
    \end{algorithmic}
\end{algorithm}

\paragraph{Variable-length Clusters}
Next, we introduce \emph{variable-length} clustering, where each cluster can contain a different number of rows depending on the similarity among consecutive rows,
for matrices where the sparsity 
structure is not repeated evenly.  
To determine where the cluster boundaries should be, inspired by the previous study on hierarchical clustering~\cite{JiangHoAg20}, we use Jaccard similarity~\cite{jaccard1901etude}, a common measure for set similarity, to measure similarity between rows. Variable-length clustering 
introduces a small computational overhead compared to fixed-length clustering due to computing Jaccard similarity scores between every pair of consecutive rows. However, it significantly improves memory efficiency with more accurate cluster groupings.
Algorithm~\ref{alg:vlength-cluster} demonstrates how to perform variable-length clustering.

The clustering process begins by iterating over the rows of \matrixA of SpGEMM (Algo.~\ref{alg:vlength-cluster} Line 4). For each cluster, the first row is chosen as the representative row (Algo.~\ref{alg:vlength-cluster} Line 2). Consecutive rows are added to the cluster if their Jaccard similarity with the representative row exceeds a predefined threshold (Algo.~\ref{alg:vlength-cluster} Line 5-6). This ensures that only structurally similar rows are grouped together. Although comparing every new row against all existing rows in the cluster would yield more accurate clusters, the associated computational cost is prohibitive—especially relative to SpGEMM runtime. Therefore, to balance accuracy and performance, we compare only against the representative row and adopt a relatively low similarity threshold (0.3 in our experiments; \textit{jacc\_th} in Algo.~\ref{alg:vlength-cluster}), while also limiting the maximum cluster size (8 in our experiments; \textit{max\_cluster\_th} in Algo.~\ref{alg:vlength-cluster}).


We illustrate this approach in ~\figref{cluster_noorder}(b). Initially, row 0 serves as the representative. As we iterate through the matrix, row 1 and row 2 have Jaccard similarities of 0.5 and 0.5, respectively, with row 0, and are added to the cluster. Row 3, however, has a similarity of 0.0, which breaks the threshold, ending the cluster at row 2. Row 3 then becomes the representative of a new cluster (Line 6 in Algorithm~\ref{alg:cluster-spgemm}). The process continues, forming clusters based on the similarity between the representative and subsequent rows: row 4 has a similarity of 0.5 with row 3 and is included, while row 5 has a similarity of 0.25 and starts a new cluster. This results in clusters: rows 0–2, 3–4, and 5 in ~\figref{cluster_noorder}(b).

\begin{figure}[t]
    \centering
    \includegraphics[width=\linewidth,page=4]{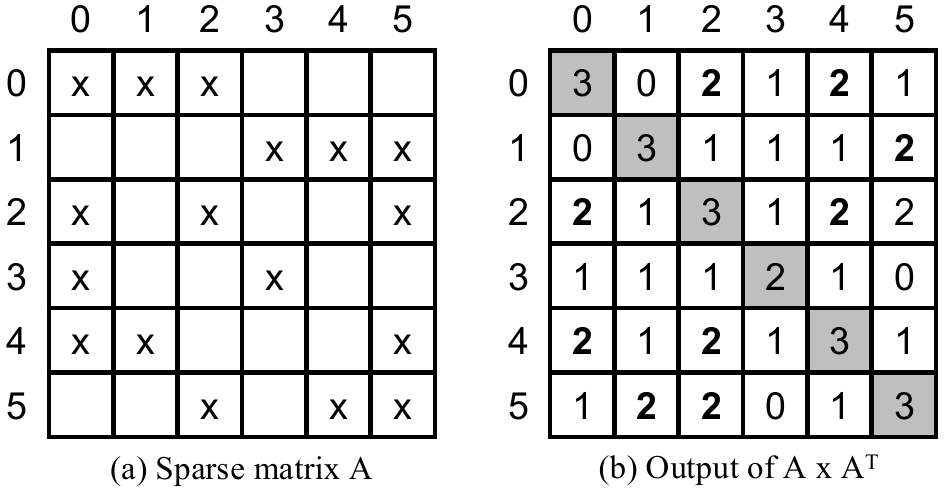}
    \caption{Hierarchical clustering.}
    \label{fig:cluster_hierarchical}
\end{figure}

\subsection{Hierarchical Clusters}
Next, we introduce a hierarchical clustering algorithm targeting SpGEMM that aims to reduce the overhead of the existing hierarchical clustering algorithm for SpMM~\cite{JiangHoAg20} without giving up quality. Hierarchical clustering targets the case where similar rows are present but not placed consecutively in the original row order of the matrix. In such cases, variable-length clustering performs suboptimally, 
creating an excessive number of small clusters. 

Although row reordering using existing algorithms
(discussed in~\secref{background}) may achieve similar objectives, it introduces high overhead and may not aim to reorder similar rows together. For example, row reordering often takes two to three orders of magnitude longer than a single SpGEMM (see~\secref{eval}). Furthermore, reordering algorithms may have different optimization objectives, such as aiming to reduce bandwidth or improve solver convergence, rather than grouping similar rows together.


We first outline how the hierarchical clustering method in~\cite{JiangHoAg20} operates, and then describe our modifications to adapt it for improved effectiveness with SpGEMM kernels. The complete procedure, including our changes, is illustrated in Algorithm~\ref{alg:hirarchical-cluster}. 

In its original form, hierarchical clustering greedily merges similar rows into clusters based on their similarity scores based on locality-sensitive hashing, which we find does not work well for SpGEMM and incurs high overhead.
We empirically observed 
that for SpGEMM, LSH-based reordering either fails to improve performance or breaks the inherent good ordering---when tested using the same parameters and similarity thresholds as in~\cite{JiangHoAg20}. Third, the LSH step itself introduces substantial overhead--- about 70  seconds on average for matrices with $10^4$ to $10^7$ rows---making it prohibitively expensive relative to the SpGEMM runtime.

To overcome these limitations, we redesign the hierarchical clustering method from~\cite{JiangHoAg20} to (1) generate candidate row pairs more efficiently via SpGEMM, and (2) construct the clustered representation based on this information, rather than only using it for reordering. Algorithm~\ref{alg:hirarchical-cluster} outlines the proposed approach.

 First, we generate candidate row pairs using a single SpGEMM computation between \matrixA and its transpose (i.e., \texttt{SpGEMM($A \times A^T$)}, Line 3). Before running this SpGEMM, we reset all \texttt{values} in \matrixA to $1$ so that the output reflects the count of overlapping nonzeros between rows---effectively capturing structural similarity.~\figref{cluster_hierarchical}(a) shows a reordered version of the \matrixA from~\figref{gustavson-spgemm}, and~\figref{cluster_hierarchical}(b) displays the output of \texttt{SpGEMM($A \times A^T$)} for this example.
Instead of storing the full output of \texttt{SpGEMM($A \times A^T$)}, we retain only the \texttt{topK} entries with the highest Jaccard similarity scores. These candidate pairs are then used in our hierarchical clustering step. 
Compared to LSH, this approach provides both faster candidate generation and more accurate similarity measurements.

The second key modification is a change in the matrix format and order of processing based on the clusters, rather than just reordering based on clusters as in prior work~\cite{JiangHoAg20}. Instead, we directly use the clusters formed via hierarchical clustering to build the \csrcluster structure for cluster-wise SpGEMM (Line 25 of Algorithm~\ref{alg:hirarchical-cluster}). 
Rather than relying on Algorithm~\ref{alg:vlength-cluster} to form clusters post-reordering, we directly adopt the cluster assignments generated by hierarchical clustering. This eliminates the need for additional similarity scans and reduces preprocessing complexity.

\begin{algorithm}[t]
    \caption{Construct Hierarchical Cluster}
    \label{alg:hirarchical-cluster}
    \begin{algorithmic}[1]
        \REQUIRE $A\_CSR[M][N], jacc\_th, max\_cluster\_th$
        \ENSURE $A\_CSR\_CLUSTER[M^*][N][K]$
        \STATE $A^T \leftarrow A.Tranpose()$
        \STATE $topk \leftarrow max\_cluster\_th - 1$
        \STATE $candidate\_pairs \leftarrow \text{SpGEMM\_TopK}(A, A^T, topk, jacc\_th)$
        \STATE \vspace{0.0em}
        \STATE $sim\_queue \leftarrow \text{MaxHeap}(candidate\_pairs)$
        \STATE $cluster\_id \leftarrow \text{Array}([0, 1, \ldots, A.nrows - 1])$
        \STATE

        \WHILE{$!sim\_queue.empty()$ \AND $nclusters > 0$}
            \STATE $(i, j) \leftarrow sim\_queue.top()$; $sim\_queue.pop()$;

            \IF{$i = cluster\_id[i]$ \AND $j = cluster\_id[j]$}
                \STATE $clusters[i], clusters[j] \leftarrow Union(i, j)$
            \ELSE
                \STATE $i \leftarrow Find(i)$
                \STATE $j \leftarrow Find(j)$
                
                \IF{$(i, j) \notin candidate\_pairs$}
                    \STATE $jacc\_score \leftarrow A\_CSR.jaccard\_similarity(i, j)$
                    \IF{$jacc\_score > jacc\_th$}
                        \STATE $sim\_queue.insert(jacc\_score, i, j)$
                        \STATE $candidate\_pairs.insert(i, j, jacc\_score)$
                    \ENDIF
                \ENDIF
            \ENDIF
        \ENDWHILE

        \STATE
        \STATE $A\_CSR\_CLUSTER(A\_CSR, clusters)$
        \RETURN $A\_CSR\_CLUSTER$
    \end{algorithmic}

\end{algorithm}

\subsection{Discussion on Cluster Building}
After presenting the various cluster-building strategies, it is important to discuss the trade-offs associated with each method. Fixed-length clustering offers the lowest cluster construction time, making it attractive for matrices with structured sparsity. However, it incurs a higher memory footprint, as it does not account for sparsity patterns and may result in significant padding. Moreover, without reordering, its effectiveness in SpGEMM performance is limited to matrices with similarly-sized groups of similar rows. 

In contrast, variable-length clustering introduces a modest overhead during cluster construction but achieves significantly better memory efficiency—even compared to the widely used compressed format, CSR. Additionally, this approach is more well-suited to a variety of sparsity patterns, offering improved performance across a broader set of matrices. 

Finally, hierarchical clustering offers a balanced trade-off between preprocessing cost and runtime improvement. Although it introduces the highest preprocessing overhead among the three methods, it delivers the best SpGEMM performance (see the boxes labeled \textit{Original} and \textit{Hierarchical} in~\figref{perf_reorder_cluster_base_rowwise}) by effectively capturing diverse sparsity patterns while maintaining a moderate memory footprint. Moreover, hierarchical clustering inherently performs row reordering during cluster formation, thereby eliminating the need for a separate reordering step.


Row reordering techniques can serve as a preprocessing step for fixed-length and variable-length clustering strategies to improve their effectiveness by generating a more amenable sparsity structure. However, the overhead of such reordering can exceed the cost of SpGEMM by many orders of magnitude. In~\secref{eval}, we quantitatively compare memory footprint and preprocessing time across all three clustering strategies.

\section{Evaluation}
\label{sec:eval}

In this section, we evaluate the impact of \numreorderings reordering algorithms, both with and without downstream clustering, across \nummatrices sparse matrices and two SpGEMM workloads. 

\subsection{Evaluation Setup}
\label{sec:eval:setup}

We developed all the SpGEMM code using C++. We used OpenMP for parallelization and Intel C++ Compiler (icpc) ver2024.1.0 with optimization level -O3. The code is publicly available on Github\footnote{https://github.com/PASSIONLab/clusterwise-spgemm}. In this section, we compare cluster-wise SpGEMM with row-wise SpGEMM and demonstrate the impact of reordering on SpGEMM performance for real-world sparse matrices. We used hashtable~\cite{nagasaka_2018_high-performance} as the sparse accumulator in all the SpGEMM experiments and 
report the average of 10 runs.

\textbf{Evaluation Platform.}
We run all our experiments on the Perlmutter supercomputer at NERSC. Perlmutter CPU nodes have two AMD EPYC 7763 (Milan) CPUs and 512 GB of DDR4 memory. Each CPU has 64 cores with 204.8 GB/s memory bandwidth, 64 MiB L2 cache and 512 GB of DDR4 memory in total. We ran all experiments on 64 threads in a single CPU.

\textbf{Datasets.}
In our evaluation, we used 110 datasets from the SuiteSparse Matrix Collection~\cite{Davis2011university}, including 26 matrices from~\cite{nagasaka_2018_high-performance} and 32 from~\cite{balaji2023community}. For the remaining datasets, we applied this selection criteria: (i) only square matrices with more than 8 million nonzeros to ensure the entire matrix do not fit in the L2 cache of our evaluation platform; (ii) matrices with less than 10 billion nonzeros due to memory limitations; and (iii) to reduce redundancy, only selected the largest matrix from each publisher-defined group, as grouped matrices often share similar characteristics~\cite{balaji2023community}. Exceptions were made for the SNAP and DIMACS10 groups, where all matrices were included due to their diverse origins~\cite{balaji2023community}.


\textbf{Workloads.}
Our evaluation tests two common workloads for SpGEMM: squaring a sparse matrix (i.e., $A^2$), and square times tall-skinny matrix~\cite{nagasaka_2018_high-performance}.~\secreftwo{clustering-perf}{reordering-perf} focus on $A^2$, while~\secref{tall-skinny} evaluates multiplication of $A$ with a tall-skinny matrix.
Several graph algorithms require executing multiple breadth-first searches (BFSs) simultaneously—for example, Betweenness Centrality (BC), which can be expressed by the multiplication of a square sparse matrix by a tall-skinny matrix. The square matrix represents the graph structure, while each column of the tall-skinny matrix represents a distinct BFS frontier, collectively forming a series of frontiers. In our experiments, we generate the tall-skinny matrices from BFS frontiers produced by CombBLAS~\cite{bulucc2011combinatorial} during BC computations. As the number of frontiers varies among datasets, we only take the first 10 forward frontier matrices in~\secref{tall-skinny}.


\subsection{Impact of Clustering on SpGEMM Performance}
\label{sec:clustering-perf}



\begin{figure}[t]
	\centering
\includegraphics[width=\linewidth]{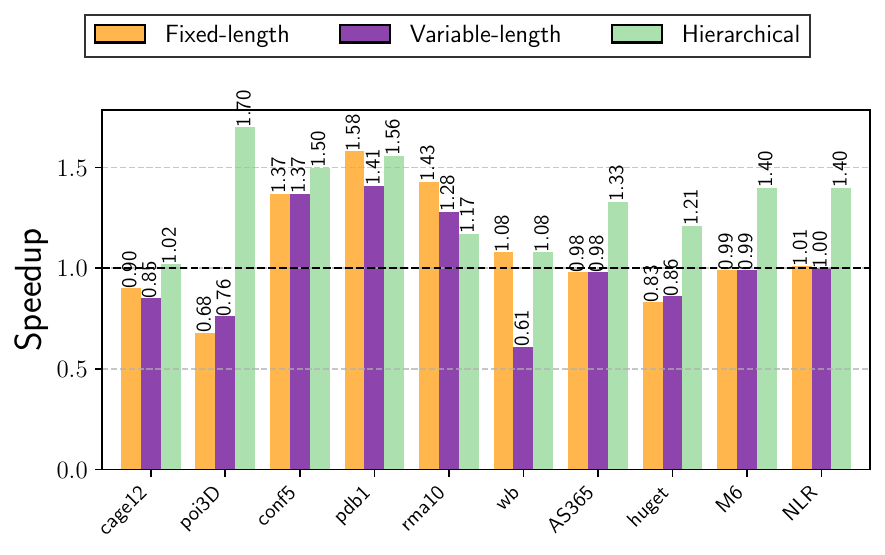}
        \vspace{-.7cm}
    \caption{\small Cluster-wise SpGEMM performance across selected datasets, relative to the row-wise SpGEMM on original matrix order.}
\label{fig:perf_clusterwise_selective}
    \vspace{.4cm}
\end{figure}

\figref{perf_reorder_cluster_base_rowwise} (on Page 2) compares the speedup of various cluster-wise SpGEMM strategies---both with and without reordering---against the traditional row-wise SpGEMM baseline across \nummatrices test datasets.
For fixed-length and variable-length methods, the results corresponding to the \texttt{Original} matrices (i.e., without reordering) are shown explicitly, while hierarchical clustering is treated as a special case under variable-length clustering. 
The y-axis represents speedup on a logarithmic scale, where values greater than $1$ indicate improved performance relative to the row-wise baseline.



Among these methods, hierarchical clustering consistently demonstrates the most
substantial performance enhancements, characterized by both a higher geomean
speedup (\meanhcspeedup) and a greater proportion of matrices with positive
outcomes—approximately \percentinputshcspeedup of matrices show performance
improvement. Fixed-length and variable-length clustering also yields positive
speedups, with approximately $45\%$ and $40\%$ of matrices without reordering
(marked as \texttt{Original}). When considering only the positive cases (i.e.,
matrices that benefit from cluster-wise SpGEMM), hierarchical clustering
achieves the highest average speedup of approximately $1.7\times$, whereas
fixed-length and variable-length clustering achieve average speedups of
approximately $1.45\times$ and $1.5\times$, respectively.

\figref{perf_clusterwise_selective} compares the three different cluster-wise SpGEMM methods on 10 representative datasets drawn from various problem domains. As shown, fixed-length and variable-length clustering strategies can improve SpGEMM performance by up to $1.58\times$ on well-structured matrices. In contrast, hierarchical clustering consistently improves performance across all 10 datasets, achieving gains of up to $1.70\times$. Hierarchical clustering achieves this superior performance through the inherent reordering of rows during the cluster construction process.

The clustering methods expose a tradeoff between preprocessing time and SpGEMM performance improvement. Hierarchical clustering introduces overhead to find similar row pairs. On the other hand, both fixed-length and variable-length clustering incur negligible preprocessing overhead. The performance of these clustering methods can be further improved by applying matrix reordering, as we will demonstrate later in this section. While reordering itself incurs a cost, this overhead can often be amortized over multiple consecutive SpGEMM executions—a common scenario in real-world applications where the same \matrixA is reused.

For example, in betweenness centrality (BC) computations, SpGEMM is executed tens of thousands of times to approximate centrality scores accurately. With a $5\%$ sampling rate on a graph containing $20$ million vertices, approximately one million breadth-first searches are required, resulting in $\mathcal{O}(1000 \times \mathit{graph\_diameter})$ SpGEMM invocations, even when using a batch size of 1000 per BC iteration~\cite{hong2024sparsity}. This makes cluster-wise SpGEMM particularly well suited for such real-world scenarios.

\begin{table}[t]
    \centering
    \caption{\small Summary of SpGEMM performance improvements achieved through reordering across different SpGEMM variants.}
    \label{tab:reorder_summary}
    \resizebox{\columnwidth}{!}{
    \begin{tabular}{l ccc ccc ccc}
    \hline
    \textbf{Algorithm} & \multicolumn{3}{c}{\textbf{Row-wise}} & \multicolumn{3}{c}{\textbf{Fixed- Cluster}} & \multicolumn{3}{c}{\textbf{Variable- Cluster}} \\
    \cline{2-10}
     & \textbf{GM} & \textbf{Pos.\%} & \textbf{+GM}
     & \textbf{GM} & \textbf{Pos.\%} & \textbf{+GM}
     & \textbf{GM} & \textbf{Pos.\%} & \textbf{+GM} \\
    \hline
    Shuffled   & 0.43   & 18.52     & 1.80      & 0.32      & 12.26     & 1.68      & 0.38      & 18.35     & 1.66 \\
    Rabbit     & 0.72   & 25.93     & 1.64      & 0.55      & 16.98     & 1.56      & 0.61      & 19.27     & 1.66 \\
    AMD        & 0.91   & 33.33     & 1.56      & 0.75      & 23.58     & 1.47      & 0.82      & 26.61     & 1.48 \\
    RCM        & 1.44   & 65.74     & 1.93      & 1.55      & 71.70     & 2.00      & 1.40      & 67.89     & 1.81 \\
    ND         & 1.33   & 57.41     & 2.09      & 1.24      & 54.72     & 1.97      & 1.17      & 52.29     & 1.92 \\
    GP         & 1.50   & 75.93     & 1.81      & 1.41      & 72.64     & 1.68      & 1.37      & 68.81     & 1.66 \\
    HP         & \B1.77 & \B79.63   & 2.14      & \B1.56    & \B80.19   & 1.80      & \B1.47    & \B76.15   & 1.76 \\
    Gray       & 1.56   & 54.63     & \B3.29    & 1.21      & 49.06     & \B2.58    & 1.34      & 45.87     & \B3.03 \\
    Degree     & 1.20   & 61.11     & 1.64      & 0.98      & 42.45     & 1.60      & 1.03      & 50.46     & 1.49 \\
    SlashBurn  & 0.91   & 36.11     & 1.46      & 0.75      & 22.64     & 1.54      & 0.81      & 32.11     & 1.38 \\
    \hline
    Best Reord. & 2.90 & 95.37 & 3.09 & 2.39 & 93.40 & 2.55 & 2.35 & 90.83 & 2.56 \\
    \hline
    \end{tabular}}
    \vspace{0.2cm}
\end{table}

\subsection{Impact of Reordering on SpGEMM Performance}\label{sec:reordering-perf}
We evaluate the impact of applying different reordering algorithms on the performance of both row-wise and cluster-wise SpGEMM across our datasets. Table~\ref{tab:reorder_summary} summarizes these results, showing geometric mean speedup (labeled as GM), portion of datasets that show positive performance improvement through reordering (labeled as Pos.\%), and the geometric mean by only considering the positive cases (labeled as $+$GM). Performance improvements are measured relative to the original ordering of matrices, where speedup values greater than $1.0$ mean better performance. The last row of Table~\ref{tab:reorder_summary} lists the best performance achievable through the reordering.

\paragraph{Reordering on Rowwise SpGEMM}
\figref{perf_reorder_rowwise} presents a performance analysis comparing the
speedup of various reordering strategies on row-wise SpGEMM compared to the
original order of the matrix across 110 test
datasets.

\begin{figure}[t]
	\centering
	\includegraphics[width=\linewidth]{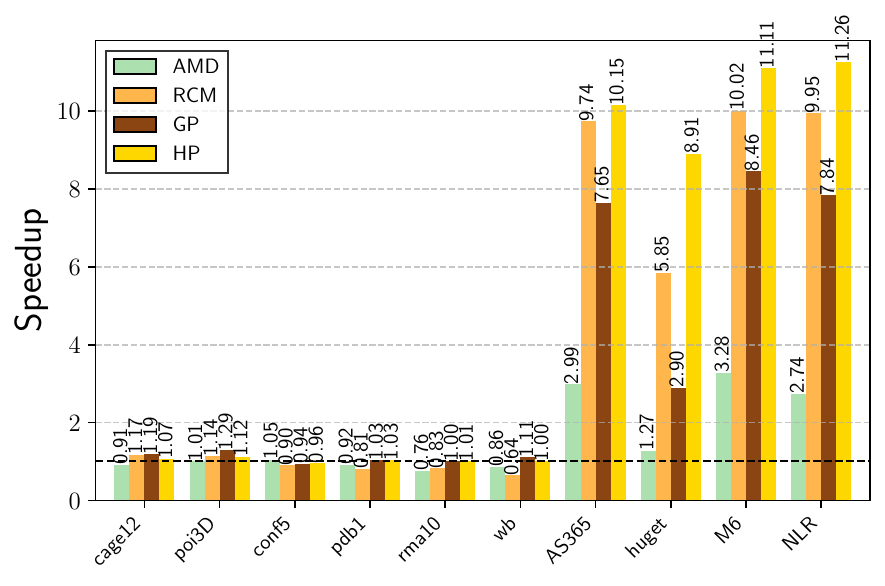}
    \caption{\small Row-wise SpGEMM performance across selected datasets and reordering algorithms, relative to the original matrix order.}
	\label{fig:perf_reorder_rowwise_selective}
\end{figure}

The results show significant variability in performance improvement across
different algorithms. Notably, HP demonstrates the highest percentage
($\approx79.6\%$) of datasets achieving positive speedups, alongside the highest
geometric mean speedup (i.e., 1.77). GP and RCM algorithms also perform effectively,
yielding positive speedups in $75\%$ and $65\%$ of datasets,
respectively. Given their strong performance, we further analyze their impact on the same ten selected datasets (previously used in~\figref{perf_clusterwise_selective}) 
in~\figref{perf_reorder_rowwise_selective}. As shown, these reorderings offer limited or comparable improvements on the first six datasets, while the remaining four demonstrate substantial speedups—reaching up to $11.26\times$. This observation underscores the fact that the effectiveness of reordering in SpGEMM is closely tied to the sparsity pattern of the input matrix.

In contrast, algorithms such as Shuffled and Rabbit show limited overall improvement, with geomean speedups below one and positive speedups on fewer than $26\%$ of datasets. However, it is worth noting that Rabbit still demonstrates strong potential: in 12 out of 110 datasets, it achieves more than $2\times$ performance improvement, with a maximum speedup of $3.32\times$ on the \texttt{M6} dataset. This highlights its relevance and potential applicability on specific inputs. 


The comprehensive performance statistics for all reordering algorithms are summarized in Table~\ref{tab:reorder_summary}. While HP, GP, and RCM consistently demonstrate superior performance improvements, all reordering algorithms are able to enhance SpGEMM performance for a subset of matrices. These findings underscore the continued relevance of employing diverse reordering strategies to optimize SpGEMM execution. Prior work~\cite{nagasaka_2018_high-performance} demonstrates that SpGEMM can achieve higher flops/s on matrices with higher compression ratio ($\approx$ flops / number of non-zeroes in the output). We find that reordering can accelerate SpGEMM, even when the compression ratio remains unchanged. This highlights an important opportunity to refine the understanding of SpGEMM performance beyond the traditional focus on compression ratio~\cite{nagasaka_2018_high-performance}. Finally, the observed maximum performance gains achieved through reordering emphasize the importance of selecting an appropriate algorithm tailored to the sparsity pattern of the input matrix.




\paragraph{Reordering on Clusterwise SpGEMM}
In~\figref{perf_reorder_cluster_base_rowwise} we present the comparative performance speedup of various cluster-wise SpGEMM strategy---both with and without reordering---across 110 test datasets. Previously,~\secref{clustering-perf} focused on the case of cluster-wise computation only. In this subsection, we consider reordering as a preprocessing step before fixed-length and variable-length cluster-wise computation, as mentioned in~\secref{clusterwise}.


\begin{table*}[t]
    \centering
    \caption{\small Speedup of row-wise SpGEMM after reordering on tall-skinny matrices, relative to the original matrix order.}
    \label{tab:ts_reorder_performance}
    \resizebox{.8\textwidth}{!}{
    \begin{tabular}{lcccccccccc|c}
        \hline
        \textbf{Dataset} & \textbf{Shuffled} & \textbf{Rabbit} & \textbf{AMD} & \textbf{RCM} & \textbf{ND} & \textbf{GP} & \textbf{HP} & \textbf{Gray} & \textbf{Degree} & \textbf{SlashBurn} & \textbf{Best Reorder} \\
        \hline

        webbase-1M & 0.79 & \cellcolor[HTML]{ACE1AF}\B1.10 & \cellcolor[HTML]{ACE1AF}\B1.23 & \cellcolor[HTML]{ACE1AF}\B1.13 & 0.85 & \B0.90 & \B0.95 & \cellcolor[HTML]{ACE1AF}\B1.03 & \B0.95 & \B0.97 & 1.23 \\

        patents\_main & \cellcolor[HTML]{ACE1AF}\B1.55 & \cellcolor[HTML]{ACE1AF}\B1.59 & \cellcolor[HTML]{ACE1AF}\B1.69 & \cellcolor[HTML]{ACE1AF}\B1.05 & \cellcolor[HTML]{ACE1AF}\B1.26 & \cellcolor[HTML]{ACE1AF}\B1.24 & \cellcolor[HTML]{ACE1AF}\B1.69 & \cellcolor[HTML]{ACE1AF}\B1.04 & \cellcolor[HTML]{ACE1AF}\B1.69 & \cellcolor[HTML]{ACE1AF}\B1.50 & 1.69 \\

        AS365 & \B0.46 & \cellcolor[HTML]{ACE1AF}\B1.28 & \cellcolor[HTML]{ACE1AF}\B1.29 & \cellcolor[HTML]{ACE1AF}\B4.50 & \cellcolor[HTML]{ACE1AF}\B4.29 & \cellcolor[HTML]{ACE1AF}\B3.14 & \cellcolor[HTML]{ACE1AF}\B4.00 & \cellcolor[HTML]{ACE1AF}\B1.68 & \cellcolor[HTML]{ACE1AF}\B1.47 & \cellcolor[HTML]{ACE1AF}\B1.15 & 4.50 \\

        com-LiveJournal & \cellcolor[HTML]{ACE1AF}\B1.69 & \cellcolor[HTML]{ACE1AF}\B2.86 & \cellcolor[HTML]{ACE1AF}\B3.05 & \cellcolor[HTML]{ACE1AF}4.65 & \cellcolor[HTML]{ACE1AF}2.91 & \cellcolor[HTML]{ACE1AF}\B2.79 & \cellcolor[HTML]{ACE1AF}\B3.01 & \cellcolor[HTML]{ACE1AF}\B1.27 & \cellcolor[HTML]{ACE1AF}\B4.04 & \cellcolor[HTML]{ACE1AF}\B1.48 & 4.65\\

        europe\_osm & 0.26 & 0.54 & 0.54 & \cellcolor[HTML]{ACE1AF}\B1.70 & \cellcolor[HTML]{ACE1AF}\B1.70 & 0.62 & \cellcolor[HTML]{ACE1AF}\B1.95 & 0.75 & \cellcolor[HTML]{ACE1AF}\B1.03 & 0.50 & 1.95 \\

        GAP-road & 0.19 & 0.47 & 0.49 & \cellcolor[HTML]{ACE1AF}\B1.32 & \cellcolor[HTML]{ACE1AF}\B1.91 & \cellcolor[HTML]{ACE1AF}\B1.68 & \cellcolor[HTML]{ACE1AF}\B1.60 & 0.42 & 0.57 & 0.43 & 1.91 \\

        kkt\_power & 0.33 & 0.39 & 0.38 & \cellcolor[HTML]{ACE1AF}\B1.25 & \cellcolor[HTML]{ACE1AF}\B1.36 & \cellcolor[HTML]{ACE1AF}\B1.22 & \cellcolor[HTML]{ACE1AF}\B1.21 & \B0.68 & \B0.75 & 0.41 & 1.36  \\

        M6 & \cellcolor[HTML]{ACE1AF}\B1.24 & \cellcolor[HTML]{ACE1AF}\B1.25 & \cellcolor[HTML]{ACE1AF}\B1.33 & \cellcolor[HTML]{ACE1AF}\B4.02 & \cellcolor[HTML]{ACE1AF}\B3.71 & \cellcolor[HTML]{ACE1AF}\B3.02 & \cellcolor[HTML]{ACE1AF}\B3.37 & \cellcolor[HTML]{ACE1AF}\B1.53 & \cellcolor[HTML]{ACE1AF}\B1.40 & \cellcolor[HTML]{ACE1AF}\B1.36 & 4.02 \\

        NLR & \B0.26 & \B0.75 & \B0.71 & \cellcolor[HTML]{ACE1AF}\B2.90 & \cellcolor[HTML]{ACE1AF}\B2.87 & \cellcolor[HTML]{ACE1AF}\B2.00 & \cellcolor[HTML]{ACE1AF}\B2.43 & \B0.86 & \B0.83 & \B0.78 & 2.87 \\

        wikipedia-20070206 & \cellcolor[HTML]{ACE1AF}\B1.68 & \cellcolor[HTML]{ACE1AF}\B2.17 & \cellcolor[HTML]{ACE1AF}\B2.86 & \cellcolor[HTML]{ACE1AF}\B2.27 & \cellcolor[HTML]{ACE1AF}\B2.46 & \cellcolor[HTML]{ACE1AF}\B2.95 & \cellcolor[HTML]{ACE1AF}\B2.17 & \cellcolor[HTML]{ACE1AF}\B1.05 & \cellcolor[HTML]{ACE1AF}\B3.42 & \cellcolor[HTML]{ACE1AF}\B1.53 & 3.42  \\
        \hline
    \end{tabular}
    }
\end{table*}

\begin{table}[t]
    \centering
    \caption{\small Speedup of hierarchical cluster-wise SpGEMM performance, relative to the row-wise SpGEMM. $i_*$ represents BC frontier iteration.}
    \label{tab:ts_hcluster_performance}
    \resizebox{\columnwidth}{!}{
    \begin{tabular}{lcccccccccc|c}
        \hline
        \textbf{Dataset} & \textbf{$i_1$} & \textbf{$i_2$} & \textbf{$i_3$} & \textbf{$i_4$} & \textbf{$i_5$} & \textbf{$i_6$} & \textbf{$i_7$} & \textbf{$i_8$} & \textbf{$i_9$} & \textbf{$i_{10}$} & \textbf{Mean} \\
        \hline
        webbase. & \cellcolor[HTML]{ACE1AF}1.00 & \cellcolor[HTML]{ACE1AF}1.45 & 0.60 & 0.71 & 0.62 & 0.61 & 0.69 & 0.93 & 0.80 & 0.71 & 0.81 \\

        patents\_m. & 0.35 & \cellcolor[HTML]{ACE1AF}1.11 & \cellcolor[HTML]{ACE1AF}1.33 & 0.89 & \cellcolor[HTML]{ACE1AF}1.23 & \cellcolor[HTML]{ACE1AF}1.11 & \cellcolor[HTML]{ACE1AF}1.04 & \cellcolor[HTML]{ACE1AF}1.06 & \cellcolor[HTML]{ACE1AF}1.03 & \cellcolor[HTML]{ACE1AF}1.04 & \cellcolor[HTML]{ACE1AF}1.02 \\

        \cellcolor[HTML]{ACE1AF}AS365 & \cellcolor[HTML]{ACE1AF}3.40 & \cellcolor[HTML]{ACE1AF}2.84 & \cellcolor[HTML]{ACE1AF}2.28 & \cellcolor[HTML]{ACE1AF}2.12 & \cellcolor[HTML]{ACE1AF}1.71 & \cellcolor[HTML]{ACE1AF}1.74 & \cellcolor[HTML]{ACE1AF}1.52 & \cellcolor[HTML]{ACE1AF}1.56 & \cellcolor[HTML]{ACE1AF}1.53 & \cellcolor[HTML]{ACE1AF}2.66 & \cellcolor[HTML]{ACE1AF}2.14 \\

        \cellcolor[HTML]{ACE1AF}com-LiveJ. & \cellcolor[HTML]{ACE1AF}1.38 & 0.78 & \cellcolor[HTML]{ACE1AF}1.18 & \cellcolor[HTML]{ACE1AF}1.02 & 0.92 & \cellcolor[HTML]{ACE1AF}1.00 & \cellcolor[HTML]{ACE1AF}1.14 & \cellcolor[HTML]{ACE1AF}1.17 & 0.77 & \cellcolor[HTML]{ACE1AF}1.03 & \cellcolor[HTML]{ACE1AF}1.04 \\

        europe\_osm & \cellcolor[HTML]{ACE1AF}1.09 & \cellcolor[HTML]{ACE1AF}1.08 & \cellcolor[HTML]{ACE1AF}1.09 & \cellcolor[HTML]{ACE1AF}1.07 & \cellcolor[HTML]{ACE1AF}1.07 & \cellcolor[HTML]{ACE1AF}1.17 & \cellcolor[HTML]{ACE1AF}1.15 & \cellcolor[HTML]{ACE1AF}1.16 & \cellcolor[HTML]{ACE1AF}1.09 & \cellcolor[HTML]{ACE1AF}1.13 & \cellcolor[HTML]{ACE1AF}1.11 \\

        GAP-road & \cellcolor[HTML]{ACE1AF}2.61 & \cellcolor[HTML]{ACE1AF}2.72 & \cellcolor[HTML]{ACE1AF}2.58 & \cellcolor[HTML]{ACE1AF}2.64 & \cellcolor[HTML]{ACE1AF}2.52 & \cellcolor[HTML]{ACE1AF}2.17 & \cellcolor[HTML]{ACE1AF}2.56 & \cellcolor[HTML]{ACE1AF}2.38 & \cellcolor[HTML]{ACE1AF}2.13 & \cellcolor[HTML]{ACE1AF}2.40 & \cellcolor[HTML]{ACE1AF}2.47 \\

        kkt\_power & \cellcolor[HTML]{ACE1AF}1.25 & \cellcolor[HTML]{ACE1AF}1.11 & \cellcolor[HTML]{ACE1AF}1.13 & 0.97 & 0.95 & 0.85 & 0.66 & 0.68 & 0.95 & \cellcolor[HTML]{ACE1AF}1.16 & 0.97 \\

        \cellcolor[HTML]{ACE1AF}M6 & \cellcolor[HTML]{ACE1AF}4.01 & \cellcolor[HTML]{ACE1AF}3.34 & \cellcolor[HTML]{ACE1AF}3.19 & \cellcolor[HTML]{ACE1AF}2.64 & \cellcolor[HTML]{ACE1AF}2.43 & \cellcolor[HTML]{ACE1AF}2.09 & \cellcolor[HTML]{ACE1AF}1.99 & \cellcolor[HTML]{ACE1AF}1.84 & \cellcolor[HTML]{ACE1AF}1.73 & \cellcolor[HTML]{ACE1AF}1.67 & \cellcolor[HTML]{ACE1AF}2.49 \\

        \cellcolor[HTML]{ACE1AF}NLR & \cellcolor[HTML]{ACE1AF}2.87 & \cellcolor[HTML]{ACE1AF}1.93 & 0.98 & 0.77 & 0.72 & 0.72 & 0.68 & 0.72 & 0.68 & 0.65 & \cellcolor[HTML]{ACE1AF}1.07 \\

        wikipedia. & 	0.96 & 	0.95 & 	0.73 & 	\cellcolor[HTML]{ACE1AF}1.05 & 	0.93 & 	0.72 & 	\cellcolor[HTML]{ACE1AF}1.09 & 	0.99 & 	\cellcolor[HTML]{ACE1AF}1.00 & 	\cellcolor[HTML]{ACE1AF}1.08 & 	0.95 \\
        \hline
    \end{tabular}}
\end{table}

Applying reordering can significantly enhance the performance of fixed-length and variable-length cluster-wise SpGEMM. For instance, while these clustering methods initially show limited performance benefits---yielding improvements in only $45\%$ and $40\%$ of matrices, respectively, compared to hierarchical clustering---their effectiveness increases substantially when combined with reordering. For example, applying HP as a preprocessing step before cluster formation boosts performance on approximately $80\%$ of the matrices, as shown in Table~\ref{tab:reorder_summary}. On the other hand, hierarchical clustering achieves performance gains with lower reordering overhead. 
This trade-off introduces an interesting optimization space between performance improvement and preprocessing overhead.

\figref{perf_reorder_cluster_base_rowwise} illustrates a similar trend with reordering alone: the HP, GP, and RCM reorderings consistently yield superior performance when combined with both fixed-length and variable-length clustering, achieving average speedups of approximately $1.5\times$ across $70\text{--}80\%$ of the datasets. In contrast, other reordering algorithms—such as Shuffled, Rabbit, AMD, and SlashBurn—generally provide limited overall benefit, with average speedups below one. However, when considering only the cases where these algorithms lead to performance improvements, they still demonstrate notable gains, indicating their potential value on specific problems.

The benefits of combining both reordering and clustering do not always compose, and the gain depends on the matrix. For example, on the NLR matrix, GP reordering alone improves SpGEMM by $7.84\times$ (~\figref{perf_clusterwise_selective}), while fixed-length or variable-length clustering alone do not improve performance much. However, adding clustering after GP brings the improvement down to $5.14-5.81\times$.
On the other hand, on the torso1 matrix, GP reordering alone improves SpGEMM by $1.70\times$, while fixed-length and variable-length clustering alone result in $3.22\times$ and $3.45\times$ improvement, respectively. In this case, the benefits compose: applying GP before clustering results in between $5.37\times - 6.21\times$ improvement. Future work includes characterizing which matrices can be sped up by combining both techniques and which can be improved by either reordering or clustering alone.

\subsection{SpGEMM Performance on Square $\times$ Tall-skinny matrix}\label{sec:tall-skinny}


To demonstrate the generalizability of the performance gain through reordering and clustering the \matrixA in SpGEMM, we evaluate SpGEMM on multiplication of a sparse A matrix with a tall-skinny B matrix. As mentioned earlier, SpGEMM with a tall-skinny matrix is a core subroutine in matrix-based graph operations.

Due to space limitations, we only report the performance of reordering on row-wise SpGEMM and hierarchical cluster-wise SpGEMM results on 10 representative datasets. The datasets are hand-picked ensuring a mix of different problem types and demonstrate good performance on $A^2$ SpGEMM among different reordering algorithms. Table~\ref{tab:ts_reorder_performance} presents the average speedup of row-wise SpGEMM involving square and tall-skinny matrices. In this context, the A matrix is a reordered square matrix, and the B matrix corresponds to one of the  (BC) frontier matrices, which are tall-skinny matrices. The square matrix undergoes a single reordering process. 

The green-highlighted cells in Table~\ref{tab:ts_reorder_performance} indicate instances where the combination of a specific dataset and reordering algorithm resulted in a positive speedup for the tall-skinny SpGEMM. Bolded values signify that the application of the corresponding reordering algorithm also achieved speedup in the $A^2$ SpGEMM scenario. The overlap between these green-highlighted cells and bolded values demonstrates that the performance improvements gained through reordering are consistent across different B matrices. This consistency suggests that the enhancements are primarily due to the increased proximity of rows with similar sparsity structures in the A matrix, rather than being dependent on the characteristics of the B matrix.

Table~\ref{tab:ts_hcluster_performance} presents the average speedup of hierarchical cluster-wise SpGEMM compared to the traditional row-wise SpGEMM across 10 BC frontier (tall-skinny) matrices. Datasets highlighted in green indicate favorable performance in hierarchical cluster-wise $A^2$ SpGEMM. Notably, hierarchical cluster-wise SpGEMM applied to tall-skinny matrices achieves superior speedup in most cases. This demonstrates that hierarchical cluster-wise SpGEMM is well-suited for real-world applications where clustering the A matrix once allows for efficient reuse in SpGEMM iterations.

\subsection{Overhead}

\begin{figure}[]
	\centering
	\includegraphics[width=\linewidth]{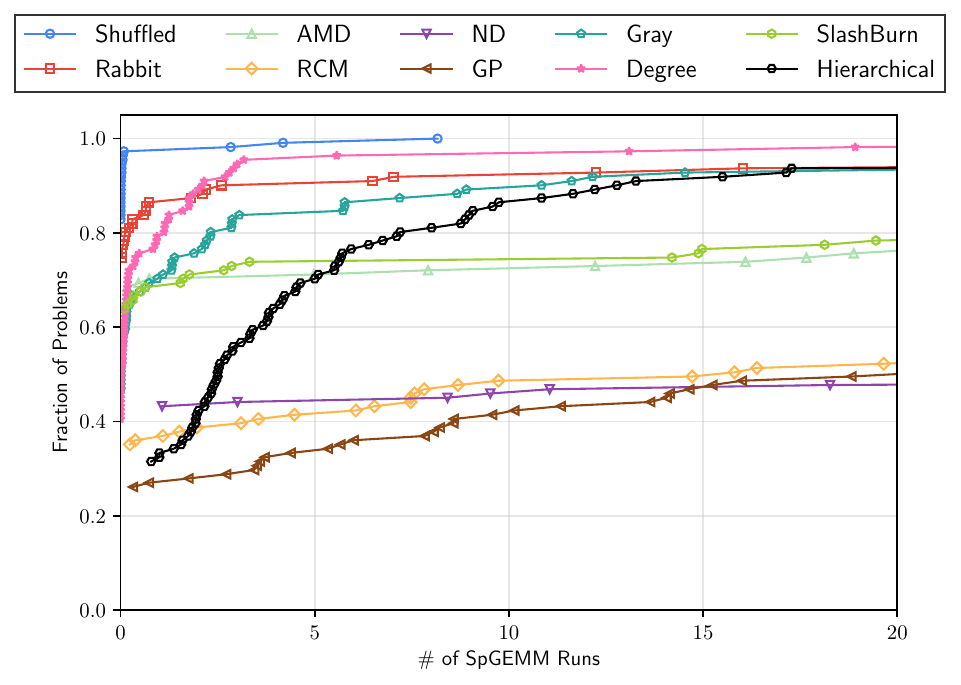}
	\caption{\small Performance profile of the overhead of reordering. For each point (x,y) in this plot, the cost of reordering is amortized after x SpGEMM iterations for y fraction of input problems.}
	\label{fig:perf_reorder_overhead}
\end{figure}

To evaluate the practicality of reordering and cluster-wise computation for SpGEMM, it is important to understand their overheads.



\paragraph{Reordering Overhead}
\figref{perf_reorder_overhead} presents the reordering overhead in terms of the number of SpGEMM iterations required to amortize the cost of reordering. This analysis considers only the cases where reordering results in performance improvement, limits the x-axis to 20 iterations for better readability, and excludes HP due to its significantly higher overhead.

The performance of various reordering algorithms in SpGEMM reveals a clear trade-off between effectiveness and overhead. Algorithms such as Shuffled, Rabbit, and Degree improve performance in a relatively small subset of datasets ($\approx10\text{--}35\%$, as shown in Table~\ref{tab:reorder_summary}), but their low reordering cost allows the overhead to be amortized quickly---within 5 SpGEMM iterations in approximately $80\%$ of cases. In contrast, RCM, GP, and HP demonstrate substantially higher effectiveness, improving performance on a broader range of datasets. However, this comes at the cost of greater reordering time, with about $50\%$ of cases requiring at least $20$ SpGEMM iterations to amortize the overhead. Gray, AMD and ND reorderings strike a moderate balance, nearly $60\%$ of the cases requiring over $100$ iterations for amortization.


In comparison, hierarchical clustering offers a more balanced trade-off. It improves SpGEMM performance in approximately $70\%$ of datasets, while $90\%$ of those cases require no more than $20$ SpGEMM iterations to amortize the clustering overhead. This level of efficiency is generally acceptable in real-world scenarios involving repeated multiplications, positioning hierarchical clustering as a practical and effective alternative.

\begin{figure}[]
	\centering
	\includegraphics[width=\linewidth]{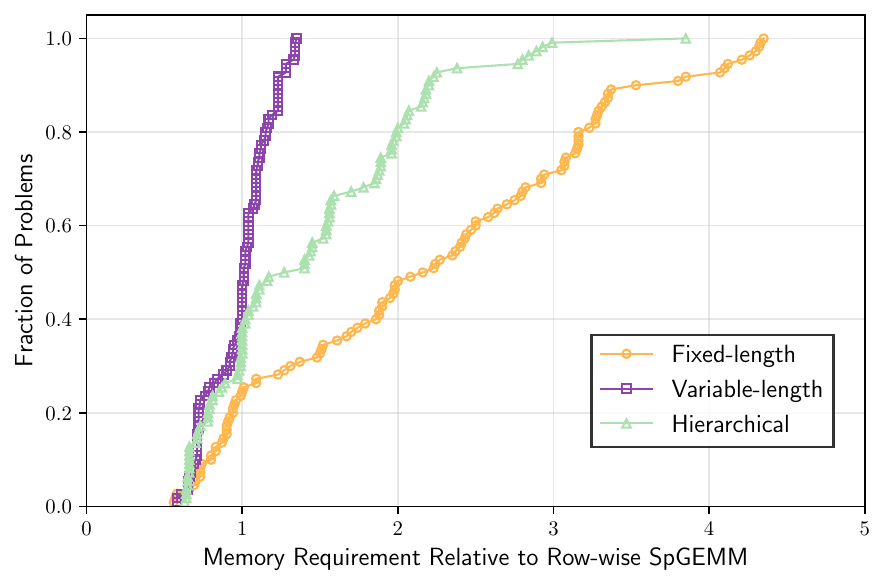}
	\caption{\small Memory overhead in cluster-wise SpGEMM relative to row-wise SpGEMM.}
	\label{fig:perf_clustering_overhead}
\end{figure}

\paragraph{Cluster-wise SpGEMM Overhead}
As discussed in~\secref{accesspattern}, the \csrcluster format may introduce space overhead. To further quantify this, we compare the memory requirements of different cluster-wise methods against the baseline row-wise approach, as shown in \figref{perf_clustering_overhead}. In this figure, each point $(X, Y)$ represents that a fraction $Y$ of the input matrices require $X\times$ memory when using cluster-wise SpGEMM compared to the row-wise baseline. Values less than $1$ on the X-axis indicate cases where \texttt{CSR\_Cluster} consumes less memory than the standard CSR format.

As the results show, variable-length clustering consistently incurs the lowest memory overhead among the three methods. In contrast, fixed-length clustering tends to require more memory, as it does not account for the sparsity pattern when forming clusters, resulting in increased padding. Hierarchical clustering strikes a balance between these two, benefiting from reordering similar rows closer together to improve memory efficiency.

Interestingly, in many cases, all three clustering strategies—fixed-length, variable-length, and hierarchical—demonstrate lower memory footprints compared to the baseline CSR format. This is because, in CSR, each nonzero value must be stored alongside its column index. In contrast, \texttt{CSR\_Cluster} groups values by column across multiple rows within a cluster, reducing the number of stored column indices and, consequently, the overall memory footprint.



%

\section{Conclusion and Future Work}
\label{sec:conclusion}

This paper introduces hierarchical clustering for SpGEMM, which combines
reordering and cluster-wise computation to improve performance by between
\rangehcspeedup on most inputs (\meanhcspeedup on average) with low
preprocessing overhead (less than \hcoverheadfactor the cost of a single SpGEMM
on about \hcoverheadproblems of inputs). Additionally, to fully characterize the
benefits of both reordering and clustering, this paper performs a comprehensive
empirical evaluation of the effect of matrix reordering and clustering, both
independently and together, on SpGEMM. Specifically, we experiment with
\numreorderings reordering algorithms and \numclusterings clustering methods on
a suite of \nummatrices matrices. To our knowledge, this is the first extensive
study of reordering algorithms in the context of SpGEMM. Overall, this paper
sheds light on the role of row reordering for SpGEMM and illustrates a tradeoff
between SpGEMM performance improvement and preprocessing time.  Future work
includes using machine learning to predict the best choice of reordering
combined with the best clustering scheme, exploring reordering for alternative
SpGEMM schemes (e.g., based on tiling), and extending the study to GPUs.
\balance


\section*{Acknowledgments}
This research is supported by the Applied Mathematics program of the Advanced Scientific Computing Research (ASCR) within the Office of Science of the DOE under Award Number DE-AC02-05CH11231. 
We used resources of the National Energy Research Scientific Computing Center (NERSC), a Department of Energy Office of Science User Facility using NERSC award ASCR-ERCAP-33069.

\bibliographystyle{ACM-Reference-Format}
{\small
\bibliography{bib}
}

\newpage

\end{document}